\documentclass[article]{revtex4}
\usepackage{amsmath}
\usepackage{graphicx}

\def\xicr{\hat{\xi}_{\rm cr}}
\def\la{\lesssim}
\def\ga{\gtrsim}

\begin{document}
\title{Constraints on cosmic ray and PeV neutrino production in blazars}
\author{B. Theodore Zhang and Zhuo Li}
\affiliation{Department of Astronomy, School of Physics, Peking
University, Beijing 100871, China\\
Kavli Institute for Astronomy and Astrophysics, Peking University,
Beijing 100871, China}

\begin{abstract}
IceCube has detected a cumulative flux of PeV neutrinos, which
origin is unknown. Blazars, active galactic nuclei with relativistic
jets pointing to us, are long and widely expected to be one of the
strong candidates of high energy neutrino sources. The neutrino
production depends strongly on the cosmic ray power of blazar jets,
which is largely unknown. The recent null results in stacking
searches of neutrinos for several blazar samples by IceCube put
upper limits on the neutrino fluxes from these blazars. Here we
compute the cosmic ray power and PeV neutrino flux of Fermi-LAT
blazars, and find that the upper limits for known blazar sources
give stringent constraint on the cosmic ray loading factor of blazar
jets (i.e., the ratio of the cosmic ray to bolometric radiation
luminosity of blazar jets), $\xi_{\rm cr}\la(2-10)\zeta^{-1}$ (with
$\zeta\la1$ the remained fraction of cosmic ray energy when
propagate into the blazar broad line region) for flat cosmic ray
spectrum, and that the cumulative PeV neutrino flux contributed by
all-sky blazars is a fraction $\la(10-50)\%$ of the IceCube detected
flux.
\end{abstract}

\maketitle

\section{Introduction}
The first discovery of high energy cosmic neutrinos had been
reported by IceCube collaboration
\citep{2013PhRvL.111b1103A,2013Sci...342E...1I,2014PhRvL.113j1101A}.
In the latest result, IceCube reported the detection of 54 events
beyond tens of TeV in the 4-yr search of high energy starting events
(HESE) in the southern hemisphere, within which three are PeV energy
scale, and the derived diffuse neutrino intensity for a flat
spectrum from 60 TeV to 3 PeV is \citep{2015arXiv151005223T}
\begin{equation}\label{eq:IC diffuse}
E_\nu^2I_\nu = 0.84 \pm 0.3 \times 10^{-8} \rm \ GeV \ cm^{-2} \
s^{-1} \ sr^{-1}
\end{equation}
The search for muon track events from
the northern hemisphere resulted in a similar intensity above 100 TeV
\citep{2015PhRvL.115h1102A}. A track event from the northern sky
with multi-PeV energy is even reported recently
\citep{2015ATel.7856....1S}.
\par
The origin of these high energy neutrinos is unknown, but believed
to be related to cosmic ray (CR) production. Astrophysical
accelerators in the universe produce the CRs which interact with
radiation ($p\gamma$) and/or matter ($pp$) in the sources or during
propagation after escape from the sources, and generate charged
pions, which decay into neutrinos via $\pi\rightarrow\nu_\mu\mu$ and
$\mu\rightarrow e\nu_e\nu_\mu$. In these processes, each neutrino
carries $\sim1/20Z$ energy of the primary CR, where $Z$ is the
charge number of the CR nuclei. The detection of PeV neutrinos
implies that there should be some kinds of cosmic accelerators that
can produce CRs up to $\sim20Z$~PeV.

\par
The arrival directions of the IceCube detected neutrinos are
consistent with isotropic distribution \citep{2014PhRvL.113j1101A}.
The gamma-ray observations of diffuse Galactic emission and
individual Galactic point sources by Fermi-LAT suggest that Galactic
neutrino flux cannot account for the total flux IceCube detects
\citep{wang14}. These hint that the IceCube neutrinos are
extragalactic origin, although some fraction could be Galactic
origin \citep[e.g.][]{neronov_MW}. There are already many
extragalactic objects proposed to produce CRs and neutrinos long
before the IceCube detection
\citep[e.g.,][]{1991PhRvL..66.2697S,waxman97,2001PhRvL..87v1102A,loeb06,2008ApJ...689L.105M,2009ApJ...707..370K}.
However, using the Fermi-LAT observations of those potential sources
and simply assuming the gamma-ray to neutrino flux ratio is
constant, \cite{wang14,2016SCPMA..59a5759W} suggested that gamma-ray
bursts and flat spectrum radio quasars (FSRQs) may not account for
the IceCube neutrino flux, but starburst galaxies are still
possible.

Blazars are a class of strong candidate sources. They are the
active galactic nuclei with relativistic jets pointing toward us, and consist of two subclasses, BL Lac objects (BL Lacs) and FSRQs. It is
expected that the CRs produced by the jet interact with photons from
or outside of the jet can produce pions and hence neutrinos.
Many discussions are going on after the IceCube detection for
whether blazars can account for the IceCube neutrinos
\citep{2014PhRvD..90b3007M,2015MNRAS.452.1877P,2015MNRAS.448.2412P,2016MNRAS.457.3582P,2016arXiv160202012K,2016arXiv160306954P,2016SCPMA..59a5759W}.

\par
Besides the detection of diffuse neutrinos, IceCube has also carried
out deep search of neutrinos from individual astrophysical objects,
including three groups of bright blazars
\citep{2014ApJ...796..109A,Schatto2014}. No signal is found but
strong upper limit is put on the neutrino flux from individual
blazars. In this work we use the flux upper limit to constrain the
CR loading in blazar jets. Moreover, applying the CR loading factor
to the population of blazars, which has been measured by Fermi-LAT
\citep[e.g.,][]{2012ApJ...751..108A,2014ApJ...780...73A}, we obtain
constraints on the all-sky neutrino flux from blazars. The results
are compared with IceCube detected flux in order to test if blazars
can produce the bulk of the IceCube PeV neutrinos. Hereafter we take
the convention $Q=10^xQ_x$ and use cgs units unless state otherwise.

\section{Neutrino production processes}
The blazar neutrino production models have been well established in
the literatures. We adopt the same picture for neutrino production
in blazars as Ref. \cite{2014PhRvD..90b3007M}, which is one of the
latest updated models. CRs are produced in the blazar jets, and
interact with the intense soft-photon field. There could be various
sources of target photons; they may come from the jet, e.g., the
synchrotron radiation from jet-accelerated electrons, whereas they
may be the broad line emission from the broad line region (BLR) or
the infrared emission from the dust torus. The neutrino production
rate depends on the CR production of the jets and the photomeson
production efficiency of CRs, which we will describe below.

We assume a connection between the CR and bolometric radiation
luminosity  from the jet as \cite{2014PhRvD..90b3007M}. Since we are
interested in the neutrino flux at $E_\nu\sim1$~PeV, we focus on the
CR production at $E_p\sim20E_\nu\sim20$~PeV (the factor 20 comes
from that the photo-produced pion carries a fraction of $\sim1/5$ of
the primary proton energy, and that each lepton from pion decay carries
$\sim1/4$ of the pion energy). We define the CR luminosity from the
jet at $E_p$ as $ E_p L_{E_p}  \approx \xicr L_{\rm rad} $, where
$L_{\rm rad}$ is the bolometric radiation luminosity of the jet, and
$\xicr$ is  the CR loading factor for CRs at $E_p\simeq20$PeV. Ref
\cite{2014PhRvD..90b3007M} defines the total CR loading factor as
$\xi_{\rm cr}\equiv L_{\rm cr}/L_{\rm rad}$ where $L_{\rm cr}$ is
the total CR luminosity. If the CR energy distribution is
$dn/dE_p\propto E_p^{-p}$ for $E_{p,\min}<E_p<E_{p,\max}$ with
$p\simeq2$, then $\xicr\simeq\xi_{\rm
cr}/\ln(E_{p,\max}/E_{p,\min})\simeq\xi_{\rm cr}/27.6$, where the
last equation is for $E_{p,\max}\sim1$~ZeV, and
$E_{p,\min}\sim1$~GeV. We assume the CR loading factor to be
constant, and the case of deviation from constant will be discussed
in the last section.

The flavor ratio of neutrinos produced by charged pion decay is
$\nu_e:\nu_\mu:\nu_\tau=1:2:0$. Due to mixing in propagation, the
observed neutrinos show roughly equal flavor ratio of
$\nu_e:\nu_\mu:\nu_\tau\approx1:1:1$. The PeV per-flavor neutrino
luminosity from the jet is given by
\begin{equation}\label{eq:nu_luminosity}
E_\nu L_{E_\nu} \approx \frac{1}{8} f_{p\gamma} E_p L_{E_p}\approx
\frac{1}{8} f_{p\gamma}\xicr L_{\rm rad}
\end{equation}
where $f_{p\gamma}$ is the photomeson production efficiency, and the
factor $1/8$ comes from that half of the primary protons are
channeled into charged pion production, and the four leptons from
charged pion decay share equally the pion energy.

Because of the photo-produced $\Delta$ resonance, the protons with
energy $E_p'$ mainly interact with photons with energy that
satisfies $E'_pE'_\gamma\approx0.2\rm GeV^2$, where both $E_p'$ and
$E_\gamma'$ should be measured in the frame where photons are
roughly isotropic. Therefore, target photons relevant to PeV
neutrino production are mainly from the broad line emission and the
synchrotron emission from the jet, other than the infrared photons
from the dust torus (which are more relevant to EeV neutrino
production). We derive the contributions to the photomeson
production efficiency by  the synchrotron emission and the broad
line emission separately.

\subsection{Synchrotron photon induced neutrino production}
The jet can accelerate electrons which then generate synchrotron and
inverse-Compton photons. In the same time, CR ions may also be
accelerated in the jet and interact with the synchrotron photons.
The CR and synchrotron photon interactions will happen unavoidably.
Consider the interactions in the comoving frame of a blob of the
jet. Given the observed jet radiation spectrum $EL_E$, the bulk
Lorentz factor of the jet $\Gamma_j$, and the blob radius $r_b$, the
comoving photon density, at comoving photon energy
$\epsilon=E/\Gamma_j$, can be given by $\epsilon n_\epsilon \approx
3E L_{E}/4\pi r_b^2 c \Gamma_j E$.

The photomeson production efficiency in the blob, for CRs
interacting with photons with energy $\epsilon$, is $
f_{p\gamma}^{\rm syn} \approx \epsilon n_{\epsilon} \sigma^{\rm
eff}_{p\gamma} \ l_b $, where $\sigma^{\rm eff}_{p\gamma}$ is the
effective cross section, and $l_b$ is the size of the blob in the
comoving frame. Using the rectangular approximation to the $p\gamma$
cross section, we have $\sigma_{p\gamma}^{\rm eff} \approx
\kappa_{\Delta} \sigma_{\Delta} (\Delta
\bar{\epsilon}_{\Delta}/\bar{\epsilon}_{\Delta})$, with
$\sigma_{\Delta} \approx 5 \times 10^{-28} {\rm cm^2}$,
$\kappa_\Delta \approx 0.2$, $\bar{\epsilon}_{\Delta} \approx 0.34
{\rm GeV}$, and $\Delta \bar{\epsilon}_\Delta \approx 0.2 {\rm
GeV}$. The typical dissipation radius is estimated to be
$r_b\approx\Gamma_jl_b$, with $l_b \approx \Gamma_j c \delta t$.
Here $\delta t$ is the variability time in the black-hole frame.
Hereafter $\Gamma_j = 10$ and $\delta t = 10^5$s are used.

For the $\sim1$~PeV neutrinos that we are interested in, the
relevant target photon energy is $E\sim1\Gamma_{j,1}^2(1\,{\rm
PeV}/E_{\nu})$keV, thus we need to know the radiation luminosity
$EL_E$ at $E=1$keV, and then the photomeson production efficiency
for CRs responsible to PeV neutrinos is
\begin{equation}\label{eq:blazarzone}
 f_{p \gamma}^{\rm syn} = 9.7 \times 10^{-6} \frac{EL_E({\rm keV})}{10^{45}\rm erg\,s^{-1}}\Gamma_{j,1}^{-4}\delta t_5^{-1}.
\end{equation}

\subsection{BLR photon induced neutrino production}
The broad line emission originates from numerous small, cold and
dense gas clouds, which are photoionized by the UV and X-rays from
the accretion disk and hot plasma. The broad line emission mainly
comes from HI $\rm  Ly\alpha$, and has photon energy of $E_{\rm
BL}=10.2$ eV in the black-hole rest frame. The corresponding
neutrino energy, due to photo-produced $\Delta$ resonance, is just
$E_\nu\sim1(10{\rm eV}/E_{\rm BL})$PeV. The typical radius of the
BLR is estimated to be $r_{\rm BLR} \approx 10^{17} \ {\rm cm} \
L^{1/2}_{\rm AD, 45}$ \citep{2008MNRAS.387.1669G}, where $L_{\rm
AD}$ is the accretion-disk luminosity. The BLR luminosity is related
to the accretion-disk luminosity through $L_{\rm BL} \approx f_{\rm
cov} L_{\rm AD}$, where $f_{\rm cov}$ is the covering factor of the
gas clouds, and typically $f_{\rm cov} = 0.1$
\citep{fcover1,fcover2}. The broad line photon density in the
black-hole rest frame is $n_{\rm BL} \approx L_{\rm BL}/4 \pi
r^2_{\rm BLR} c E_{\rm BL}$.

The CRs produced inside the BLR will propagate through the BLR, and
unavoidably interact with the broad line photons and generate pions.
The photomeson production efficiency in this process, in which PeV
neutrinos are produced, is then $ f_{p\gamma}^{\rm BLR} \approx
n_{\rm BL} \sigma^{\rm eff}_{p\gamma} r_{\rm BLR} $, i.e.,
\begin{equation}\label{eq:blr}
  f_{p\gamma}^{\rm BLR}=9.6 \times 10^{-3} f_{\rm cov, -1}L_{\rm
AD, 45}^{1/2}.
\end{equation}
The CRs can interact with both synchrotron and BLR photons to
generate PeV neutrinos,  and it can be seen that usually the
efficiency induced by BLR photons is much higher than the one by
synchrotron photons, $f_{p\gamma}^{\rm BLR}\gg f_{p\gamma}^{\rm
syn}$.

The total photomeson production efficiency in
eq.(\ref{eq:nu_luminosity}) is
\begin{equation}\label{eq:total f_pgamma}
  f_{p\gamma}=f_{p\gamma}^{\rm syn}+\zeta f_{p\gamma}^{\rm BLR},
\end{equation}
where $\zeta<1$  is the fraction of the CR energy after the energy
loss before they propagate into the BLR. The energy loss could be
due to, e.g., photomeson production by interactions with synchrotron
photons, and adiabatic cooling, etc. Usually $f_{p \gamma}^{\rm
syn}\ll1$ suggests that the former is negligible, then the latter is
more relevant, i.e., $\zeta\ga r_b/r_{\rm
BLR}\approx0.3\Gamma_{j,1}^2\delta t_{5}L_{\rm AD,45}^{-1/2}$, where
the lower limit holds if the CRs are confined in the blob until the
blob travels to the BLR. So the PeV neutrino production is
dominated by CR-BLR photon interactions, $\zeta f_{p\gamma}^{\rm
BLR}\gg f_{p\gamma}^{\rm syn}$, as in Ref
\cite{2014PhRvD..90b3007M}.

\subsection{Blazar emission}
In order to derive the neutrino luminosity $E_\nu L_{E_\nu}$ for a
blazar with eqs. (\ref{eq:nu_luminosity}), (\ref{eq:blazarzone}),
(\ref{eq:blr}) and (\ref{eq:total f_pgamma}), we need its accretion
disk luminosity $L_{\rm AD}$, bolometric radiation luminosity
$L_{\rm rad}$, and specific luminosity at 1~keV $EL_E(\rm keV)$. For
this purpose we follow the blazar emission model in ref.
\cite{2014PhRvD..90b3007M}, which adopted the accretion-disk model
of \citep{2010A&A...512A..34L,1994ApJS...95....1E} and the blazar
sequence model of \cite{2011PhRvD..84j3007A,2012JCAP...11..026H}.
Specifically, in order to derive the neutrino luminosity of
Fermi-LAT detected blazars, we need $L_{\rm AD}$, $L_{\rm rad}$, and
$EL_E(\rm keV)$ as functions of $L_\gamma$, the gamma-ray luminosity
integrated over the spectrum above 100 MeV.

Ref \cite{2014PhRvD..90b3007M} gives discrete values of $L_{\rm AD}$
and $L_{\rm rad}$, and by linear fits we obtain
\begin{eqnarray}
  \log L_{\rm rad} = 1.1 x - 2.2,\nonumber\\
  \log L_{\rm AD}= 1.3 x - 16,\nonumber
\end{eqnarray}
where $x\equiv\log L_\gamma$, and the luminosities are in units of
erg s$^{-1}$. The linear relations can be expected given the facts
that gamma-ray luminosity is a significant fraction of the
bolometric radiation luminosity, and that the bolometric radiation
luminosity is roughly proportional to the accretion disk luminosity.

The keV luminosity depends on the blazar spectra, which are found to
follow the blazar sequence model. The relation of $EL_E(\rm keV)$
with $L_\gamma$ shows a weak function of $L_\gamma$ at low
$L_\gamma$, because of the fact that as the luminosity increases the
synchrotron spectral peak moves to lower frequencies, and a maximum
around $L_\gamma\sim10^{49.6}\rm erg\,s^{-1}$, due to the crossing
of the synchrotron spectral peak at keV scale. We adopt a three
segment fit to the relation, and obtain
\begin{eqnarray}
  \log EL_E({\rm keV}) = 1.3\times10^{-14}x+45 & x<46\nonumber\\
  = 0.27 x^2 - 25x+630 & 46<x<49.6\nonumber\\
  = -0.92x+93   &  x>49.6 \nonumber
\end{eqnarray}

Now for a blazar with its gamma-ray luminosity $L_\gamma$ measured,
the main uncertainty in deriving its PeV neutrino luminosity $E_\nu
L_{E_\nu}$ is the unknown CR loading factor $\xicr$. On the other
hand, once $L_\gamma$ and $E_\nu L_{E_\nu}$ are measured, $\xicr$
can be determined or constrained. Note that unlike
\cite{2016SCPMA..59a5759W} which assumes simple proportionality of
neutrino and gamma-ray luminosities, here $L_{\rm rad}$ and
$f_{p\gamma}$ are not strict linear function of $L_\gamma$, so the
neutrino luminosity $E_\nu L_{E_\nu}$ (eq. \ref{eq:nu_luminosity})
is not strictly linear function of $L_\gamma$ either.

It should be commented that the errors in the above three luminosity
relations are all smaller than 0.3 dex. Moreover there may be
uncertainty in the bulk Lorentz factor of jets $\Gamma_j$. If
$\Gamma_j=20$ then the relevant synchrotron photons for PeV neutrino
production lie at 4~keV. We will discuss below the effects from the
uncertainties of the luminosities and $\Gamma_j$.

\section{Constraints on CR loading}
IceCube had tried to search the muon track events from known
blazars. They select three catalogs of blazars, i.e., 33 bright
FSRQs, 27 low synchrotron peaked (LSP) BL Lacs, and 37 hard spectrum
BL Lacs\citep{2014ApJ...796..109A,Schatto2014}. All the searches
result in non-detection, and put constraints on their neutrino flux.
As for the 33 bright FSRQs that IceCube selected, all of them are
compatible with the null hypothesis. The 90\% confidence level upper
limit for the combined $\nu_\mu + \bar{\nu}_\mu$ flux from the 33
FSRQs is \citep{2014ApJ...796..109A,Schatto2014}
\begin{equation}\label{eq:stackinglimit}
E_\nu^2 \Phi_{\nu_\mu + \bar{\nu}_\mu}^{90\%} = 3.46 \times 10^{-9}
\rm \ {GeV \, cm^{-2}  s^{-1}}.
\end{equation}

Note that a flat neutrino spectrum, $\Phi_{\nu}\propto E_\nu^{-2}$,
is assumed in giving the upper limit. For the other assumed spectral
profiles, the upper limit for the neutrino flux at PeV will change.
For example, assume the spectral shape derived by Ref.
\cite{2014PhRvD..90b3007M}, then the spectrum is harder, with the
TeV-to-PeV slope roughly following $\Phi_{\nu}\propto
E_\nu^{-1.23}$. Note also that the IceCube upper limits are resulted
from stacking sources from both northern and southern hemispheres.
IceCube in the south pole is more sensitive to northern point
sources at TeV-to-PeV range, whereas less sensitive to southern
sources. To be conservative we assume all stacked sources come from
northern hemisphere. For track events from the northern hemisphere,
the IceCube effective area in the relevant neutrino energy range can
be approximated as $A_{\rm eff}(E_\nu)\propto E_\nu$, then the upper
limit for the neutrino flux at $E_\nu=1$~PeV is, compared with that
with flat spectrum assumed, enhanced by a factor of
$$
f_\alpha=\frac{\int_{\rm 1~TeV}^{\rm 1~PeV}(E_\nu/{\rm
1~PeV})^{-2}A_{\rm eff}(E_\nu)dE_\nu}{\int_{\rm 1~TeV}^{\rm
1~PeV}(E_\nu/{\rm 1~PeV})^{-1.23}A_{\rm eff}(E_\nu)dE_\nu}=5.3.
$$
This factor is smaller if consider the much weaker sensitivity of
IceCube for southern point sources. Unless stated otherwise, we
consider in the following the IceCube upper limits with flat
neutrino spectrum assumed, but the case of harder spectra will also
be discussed.
\par

We calculate the total neutrino flux from the 33 FSRQs, given their
gamma-ray emission properties, then the CR loading factor $\xicr$ can
be constrained by the IceCube limit on the neutrino flux. The
neutrino flux from a certain FSRQ is given by
\begin{equation}\label{eq:individual flux}
  E_\nu^2\Phi_{\nu,i}=
\frac{1}{8} f_{p\gamma}(L_{\gamma,i})\frac{ L_{\rm rad}(L_{\gamma,i})}{L_{\gamma,i}}\xicr S_{\gamma,i}
\end{equation}
where $L_{\gamma,i}$ and $S_{\gamma,i}$ are its gamma-ray luminosity
and observed flux, respectively, in the range of 0.1-100 GeV.  In
Table B.6 of \cite{Schatto2014}, the photon number flux, the
spectral photon indices and gamma-ray luminosity have been listed
for these 33 FSRQs. We take the gamma-ray luminosity from the table
(note they are calculated under the same cosmology model as we do
later in the calculation of the diffuse neutrino intensity from
blazars). We also take the gamma-ray flux from the table, but note
that the table only gives the photon number flux of 1 GeV to 100
GeV, and we convert them to the gamma-ray flux $S_\gamma$ of 100 MeV
to 100 GeV with the given spectral photon indices (resulted from
spectral fitting in 0.1-100GeV range). The interactions with BLR
photons dominate the production of PeV neutrinos for FSRQs, thus
only eq (\ref{eq:blr}) is relevant for the calculation of
$f_{p\gamma}$. Using the observational limit in stacking search, eq
(\ref{eq:stackinglimit}), we obtain an upper limit on the CR loading
factor;
\begin{equation}\label{eq:cr loading factor}
  \sum_i E_\nu^2\Phi_{\nu,i}<
E_\nu^2 \Phi_{\nu_\mu + \bar{\nu}_\mu}^{90\%} \Rightarrow
\xicr<0.062f_{\rm cov,-1}^{-1}\zeta^{-1}.
\end{equation}
As for the total CR loading factor, we have $\xi_{\rm
cr}\simeq27.6\xicr\la1.7f_{\rm cov,-1}^{-1}\zeta^{-1}$ for flat CR spectrum, $p\simeq2$.

\par
With the same method we also constrain $\xicr$ based on the upper
limit that IceCube put on the neutrino flux of the 27 LSP BL Lacs,
for which the gamma-ray properties have been shown in Table B.7 of
\cite{Schatto2014}. The 90\% confidence level upper limit for the
$\nu_\mu + \bar{\nu}_\mu$ flux is \citep{Schatto2014} $ E_\nu^2
\Phi_{\nu_\mu + \bar{\nu}_\mu}^{90\%} = 5.24 \times 10^{-9} \rm {GeV
\, cm^{-2} s^{-1}} $. In the calculation of neutrino flux from
individual sources, we consider neutrino production induced by both
BLR and synchrotron photons, but the contribution by synchrotron
photon induced production is much smaller, thus the constraint on
$\xicr$ mainly results from the BLR induced neutrinos. Our result
gives $\xicr<0.92\zeta^{-1}$, and hence $\xi_{\rm
cr}\la25\zeta^{-1}$. The constraint is much less stringent than
using FSRQs, because LSP BL Lacs are weaker neutrino producers. We
do not expect FSRQs and BL Lacs are different in CR production
mechanisms, so the CR loading factors should depend on intrinsic
physics of jets. In the following we apply the more stringent
constraint, eq (\ref{eq:cr loading factor}), to both FSRQs and BL
Lacs, and calculate their contribution to the diffuse neutrinos. It
should be noted that if assuming a harder sub-PeV neutrino spectrum
like in Ref. \cite{2014PhRvD..90b3007M}, the upper limits to the CR
loading factor (eq.\ref{eq:cr loading factor}) is increased by a
factor of $f_\alpha=5.3$, i.e., $\xicr<0.33\zeta^{-1}$ and $\xi_{\rm
cr}\la9\zeta^{-1}$.

\section{Constraints on diffuse PeV neutrinos from blazars}
Since the blazar neutrino luminosity $E_\nu
L_{E_\nu}(L_\gamma;\xicr)$ can be calculated as above given the
gamma-ray luminosity $L_\gamma$ and the CR loading factor $\xicr$,
the all-sky neutrino flux from all blazars should depend on the
redshift and gamma-ray luminosity distribution of blazars, which has
been measured recently thanks to the deep all-sky survey by
Fermi-LAT \citep{2012ApJ...751..108A,2014ApJ...780...73A}. The
observed diffuse neutrino intensity is the sum of all blazar
contribution in the universe, and is given by
\citep[e.g.,][]{2012ApJ...751..108A}
\begin{eqnarray}\label{eq:diffuse}
E_\nu^2I_\nu= \frac{c}{4\pi H_0} \int_{z_{\rm min}}^{z_{\rm max}}
\!dz\frac{1}{(1 + z)^2\sqrt{(1+z)^3 \Omega_m+\Omega_\Lambda }} \nonumber\\
\times
\int_{L_{\gamma, \rm min}}^{L_{\gamma, \rm max}} dL_{\gamma}
\int_{\Gamma_{\min}}^{\Gamma_{\max}}d\Gamma\frac{d \rho
}{d{L_\gamma}} E_\nu L_{E_\nu},
\end{eqnarray}
where $E_\nu L_{E_\nu}$ is the neutrino luminosity from one blazar,
given by eq. (\ref{eq:nu_luminosity}), and ${d \rho}/{dL_\gamma}$ is
the blazar density in the universe. Following
\cite{2012ApJ...751..108A,2014ApJ...780...73A}, here we consider ${d
\rho}(z,L_\gamma,\Gamma)/{dL_\gamma}$ as function of not only the
redshift $z$ and gamma-ray luminosity $L_\gamma$, but also the
spectral photon index $\Gamma$ in gamma-rays, thus ${d
\rho}(z,L_\gamma,\Gamma)/{dL_\gamma}$ is the blazar number per unit
comoving volume per unit luminosity and per unit $\Gamma$. For the
cosmology we take $H_0 = 67.8 \rm km\,s^{-1}Mpc^{-1}$, $\Omega_m =
0.315$, and $\Omega_\Lambda = 0.685$.

\paragraph*{FSRQs contribution.}
We consider the diffuse PeV neutrinos from FSRQs first. We adopt the
redshift and luminosity distributions of FSRQs ${d
\rho}/{d{L_\gamma}}$ from \cite{2012ApJ...751..108A}, which shows
that the luminosity-dependent density evolution (LDDE) model gives
the best fit among the others to the Fermi-LAT data. The LDDE model
is parameterized as:
$$
\frac{d\rho}{dL_\gamma}(L_\gamma,z,\Gamma)= \Phi(L_\gamma) \times
e(z, L_\gamma) \times e^{-\frac{(\Gamma - \mu)^2}{2 \sigma^2}}, $$
$$ \Phi(L_\gamma) =
\frac{A}{\ln(10)L_\gamma}\left[\left(\frac{L_\gamma}{L_*}
\right)^{\gamma_1} +
\left(\frac{L_\gamma}{L_*}\right)^{\gamma_2}\right]^{-1} ,$$
 $$   e(z, L_\gamma) = \left[\left(\frac{1+z}{1+z_c(L_\gamma)}\right)^{p_1} + \left(\frac{1+z}
{1+z_c(L_\gamma)}\right)^{p_2}\right]^{-1}, $$ with $z_c(L_\gamma) =
z_c^* (L_\gamma/10^{48})^\alpha $. Here $z_c(L_\gamma)$ corresponds
to the redshift where the evolution changes signs.
For the best fit the parameter values are: $A = 3.06\times
10^{-9}\mathrm{Mpc^{-3}}$, $\gamma_1 = 0.21 $, $L_\star =
0.84\times10^{48}\rm erg\,s^{-1}$, $\gamma_2 = 1.58$, $z_c^\star =
1.47$, $\alpha =0.21 $, $p_1 = 7.35$, $p_2 = -6.51$, $\mu = 2.44$,
and $\sigma = 0.18$.
The integration range is as follows: for the redshift range $z_{\rm
min} = 0.01$ and $z_{\rm max} = 6$; for the luminosity range
$L_{\gamma, \rm min} = 10^{44} \ \rm erg \ s^{-1}$ and $L_{\gamma,
\rm max} = 10^{52} \ \rm erg \ s^{-1}$; and for the photon index
range $\Gamma_{\rm min} = 1.8$ and $\Gamma_{\rm max} = 3.0$.
\par
Using eq (\ref{eq:cr loading factor}), the derived diffuse PeV
neutrino intensity (per flavor) from FSRQs induced by BLR photons is
\begin{equation}\label{eq:fsrq_diffuse}
E_\nu^2 I_\nu < 0.74 \times 10^{-9}  \rm GeV\,  cm^{-2}  s^{-1}
sr^{-1}
\end{equation}
(note $\zeta$ is canceled out) whereas that induced by synchrotron
photons is $ E_\nu^2 I_\nu < 1.49 \times 10^{-12}\zeta^{-1} \rm
GeV\, cm^{-2} s^{-1} sr^{-1} $. The FSRQ neutrino production is
dominated by that induced by BLR photons, but the upper limit on
their diffuse flux is only a faction $<8.8\%$ of the IceCube
detected flux (eq. \ref{eq:IC diffuse}).

\paragraph*{BL Lac contribution.}
Next consider LSP BL Lac objects. The density distribution of LSP BL
Lacs is taken from \citep{2014ApJ...780...73A}, which also shows
that the LDDE model provides a better representation of the data: $$
\frac{d\rho}{dL_\gamma}(L_\gamma,z,\Gamma) = \Phi(L_\gamma, z = 0,
\Gamma) \times e(z, L_\gamma) ,$$
$$ \Phi(L_\gamma, z = 0, \Gamma)
 = \Phi(L_\gamma)\times
e^{-\frac{(\Gamma - \mu(L_\gamma))^2}{2\sigma^2}}, $$  $$
    e(z, L_\gamma) = \left[\left(\frac{1+z}{1+z_c(L_\gamma)}\right)^{p_1(L_\gamma)} + \left(\frac{1+z}
{1+z_c(L_\gamma)}\right)^{p_2}\right]^{-1} ,$$ where $\mu(L_\gamma)
= \mu^* + \beta \times (\log(L_\gamma) - 46)$, $p_1(L_\gamma) =
p_1^* + \tau \times(\log(L_\gamma) - 46)$, and $z_c(L_\gamma) =
z_c^* (L_\gamma/10^{48})^\alpha$. The values of parameters from the
best fit are: $A = 3.34 \times10^{-10}\ \mathrm{Mpc^{-3}}$,
$\gamma_1 = 0.48$, $\gamma_2 = 6.33$, $L_* = 1.48\times 10^{48}\rm
erg\,s^{-1}$, $z_c^* = 0.96$, $\alpha = -1.73 \times 10^{-3}$,
$p_1^* = 4.10$, $\tau = 5.34$, $p_2 = -5.53$, $\mu^* = 2.32$, $\beta
= -3.24 \times 10^{-2}$, and $\sigma = 0.23$. The integration ranges
are, for redshift $z_{\rm min} = 0.03$ and $z_{\rm max} =  6$; for
luminosity $L_{\gamma, \rm min} = 7 \times 10^{43} \ \rm erg \
s^{-1}$ and $L_{\gamma, \rm max} = 10^{52} \ \rm erg \ s^{-1}$; and
for photon index $\Gamma_{\rm min} = 1.45$ and $\Gamma_{\rm max} =
2.80$. With eq (\ref{eq:cr loading factor}) the derived diffuse PeV
neutrino intensity (per flavor) from LSP BL Lacs is, for neutrinos
induced by BLR photons, $ E_\nu^2 I_\nu < 9.94 \times 10^{-11} \rm
GeV \, cm^{-2} s^{-1}
 sr^{-1}
$, whereas for that induced by synchrotron photons, $ E_\nu^2 I_\nu
< 7.98 \times 10^{-14}\zeta^{-1}  \rm GeV \, cm^{-2} s^{-1}
 sr^{-1}
$.
\par

Finally we consider the diffuse neutrinos contributed by high and
intermediate synchrotron peaked (HSP, ISP) BL Lacs. Their density
distribution is described by the same LDDE model as LSP BL Lacs, but
the best fit parameter values are different
\citep{2014ApJ...780...73A}: $A = 29.1 \times10^{-10}
\mathrm{Mpc^{-3}}$, $\gamma_1 = 0.22$, $\gamma_2 = 2.10$, $L_* =
0.26\times10^{48}\rm erg\,s^{-1}$, $z_c^* = 1.46$, $\alpha = 9.41
\times 10^{-2}$, $p_1^* = 1.98$, $\tau = 6.38$, $p_2 = -8.29$,
$\mu^* = 2.05$, $\beta = 5.55 \times 10^{-2}$, and $\sigma = 0.24$.
For the integration limits we use: $z_{\rm min} = 0.03$, $z_{\rm
max} = 6$, $L_{\gamma, \rm min} = 7\times 10^{43} \ \rm erg \
s^{-1}$, $L_{\gamma, \rm max} = 10^{52} \ \rm erg \ s^{-1}$,
$\Gamma_{\rm min} = 1.45$, and $\Gamma_{\rm max} = 2.80$. The
derived diffuse PeV neutrino flux (per flavor) induced by
synchrotron photons for HSP and ISP BL Lacs is, with eq (\ref{eq:cr
loading factor}), $ E_\nu^2 I_\nu< 2.78 \times 10^{-13}\zeta^{-1}
\rm GeV \, cm^{-2} s^{-1}
 sr^{-1}
$. We have neglected the neutrino production in BLR, because the
broad line emission is weak in HSP and ISP BL Lacs. We can see that
the neutrino flux from BL Lacs are many orders of magnitude smaller
than the IceCube detection (eq. \ref{eq:IC diffuse}).

We summarize in Table \ref{table:diffuse} the constraints on $\xicr$
for different blazar groups, and hence their contributions to the
diffuse PeV neutrino intensity.
\begin{table}[t]
\begin{center}
\caption{Upper limits for the diffuse PeV neutrino intensity
$E_\nu^2I_\nu$ from different types of blazars.
}\label{table:diffuse}
\begin{tabular}{lllll}
\tableline
Blazar type &$\xicr\zeta$ & BLR & Synchr & Fraction\\
\tableline
FSRQ & 0.062  &  $0.74 \times 10^{-9}$ &  $1.49  \times 10^{-12}\zeta^{-1}$ & 0.088 \\
LSP BL Lac & 0.92  & $1.45 \times 10^{-9}$ & $1.18 \times 10^{-12}\zeta^{-1}$  & 0.17 \\
 & 0.062   & $9.94 \times 10^{-11}$ & $7.98 \times 10^{-14}\zeta^{-1}$  & 0.012 \\
HSP+ISP BL Lac &  0.92  &  & $4.12 \times 10^{-12}\zeta^{-1}$ & $4.9\times10^{-4}\zeta^{-1}$ \\
 &  0.062  &  & $2.78 \times 10^{-13}\zeta^{-1}$ & $3.3\times10^{-5}\zeta^{-1}$ \\
\tableline
\end{tabular}
\end{center}
Notes. All intensities are in unit of $\rm GeV \, cm^{-2}  s^{-1}
sr^{-1}$. The last column is the fraction of the blazar diffuse
neutrino intensity within the IceCube detected intensity (eq.
\ref{eq:IC diffuse}).
\end{table}

Let us discuss the model uncertainties in our derivation of the
diffuse neutrino flux from blazars. (1) If assume a harder sub-PeV
neutrino spectrum, all the upper limits for the diffuse PeV neutrino
intensities should be increased by the same factor $f_{\alpha}$ as
the CR loading factor. For $f_\alpha=5.3$, the limit for the diffuse
neutrino intensity of FSRQs (eq. \ref{eq:fsrq_diffuse}) becomes
$E_\nu^2 I_\nu < 0.39 \times 10^{-8}  \rm GeV\,  cm^{-2}  s^{-1}
sr^{-1}$, only a fraction $<47\%$ of the IceCube detection.

(2) There is uncertainty in the measured blazar density distribution
$d\rho/dL_\gamma$. However, this only leads to the uncertainty of
the derived total neutrino intensity similar to that of derived
gamma-ray background intensity, i.e., a factor of
$\sim20\%$\citep{2012ApJ...751..108A,2014ApJ...780...73A}.

(3) Some parameter values have been fixed in the calculation. The
$f_{\rm cov}$ value affects the constraint on $\xicr$, but it is
canceled out in deriving the diffuse neutrino flux. The values of
$\Gamma_j$ and $\delta t$ do affect the all-sky flux of synchrotron
photon induced neutrinos, but the change should be within orders of
magnitude. For example, if $\Gamma_j=20$, the relevant photon energy
for PeV neutrinos is $E=4$~keV. We derive again the $EL_E$ and
$L_\gamma$ relation, and then calculate the synchrotron photon
induced neutrino flux $E_\nu^2I_\nu$ from FSRQs, LSP BL Lac, and
HSP+ISP BL Lac, which are changed by a factor of 2, 0.9, and 0.4,
respectively. Thus the conclusion that the diffuse PeV neutrino flux
induced by synchrotron photons is smaller than the IceCube flux by
many orders of magnitude does not change.

(4) There may be effect from the uncertainties in the relations of
$L_\gamma$ with $L_{\rm AD}$ and $L_{\rm rad}$. The constraint of
$\xicr$ with FSRQ samples does depend on determinations of $L_{\rm
AD}$ and $L_{\rm rad}$, however, as seen by comparing eqs.
(\ref{eq:cr loading factor}) and (\ref{eq:diffuse}), the dependence
is largely canceled out in deriving the diffuse intensity of BLR
induced neutrinos. For example, we have arbitrarily change $L_{\rm
rad}$ by a factor of 3 to 1/3, the upper limit on $\xicr$ (eq.
\ref{eq:cr loading factor}) varies from 0.02 to 0.18, and the upper
limit on $E_\nu^2I_\nu$ of FSRQs (eq.\ref{eq:fsrq_diffuse}) from
0.78 to $0.69 \times10^{-9}  \rm GeV\,  cm^{-2}  s^{-1} sr^{-1}$.
Also, we change $L_{\rm AD}$ by a factor of 3 to 1/3, the upper
limit on $\xicr$ (eq. \ref{eq:cr loading factor}) varies from 0.034
to 0.1, and the upper limit on $E_\nu^2I_\nu$ of FSRQs from 0.72 to
$0.71 \times10^{-9}  \rm GeV\,  cm^{-2}  s^{-1} sr^{-1}$. So the
limit on the all-sky neutrino flux induced by BLR photons is hardly
affected by the uncertainty in determining $L_{\rm AD}$ and $L_{\rm
rad}$. Note the relations of $L_{\rm AD}$ and $L_{\rm rad}$ with
$L_\gamma$ hardly depend on the adopted blazar sequence model, which
is still under debate \citep[e.g.,][]{2012MNRAS.420.2899G}.

\section{Summary and discussion}
We have used three observational results to constrain the diffuse
PeV neutrino intensity from blazars, i.e.,  (1) the constraint of
neutrino flux from known blazars in stacking searches by IceCube,
(2) the measured gamma-ray flux from these known blazars by
Fermi-LAT, and (3) the density distribution of blazars in the
universe measured by Fermi-LAT. We first use results (1) and (2) to
constrain the CR loading factor $\xicr$ of FSRQs. The result
indicates that the total CR loading is small, $\xi_{\rm cr}\equiv
L_{\rm cr}/L_{\rm rad}\la2\zeta^{-1}$ for flat CR spectrum. Given
that the physics of CR production in blazar jet is expected to be
intrinsically the same, the constraint of $\xicr$ is applied to BL
Lacs as well. With result (3) we then derive the diffuse PeV
neutrino intensity from blazars (Table \ref{table:diffuse}). We find
that blazar neutrinos, dominated by FSRQs, can only contribute
$\la10\%$ of the diffuse PeV neutrino intensity (eq. \ref{eq:IC
diffuse}). The conclusion is consistent with the result by simply
assuming constant ratio of neutrino to gamma-ray flux from blazars
\citep{2016SCPMA..59a5759W}.

If the cumulative neutrino spectrum from blazars is as hard as that
of Ref. \cite{2014PhRvD..90b3007M}, the conclusions become that the
total CR loading is $\xi_{\rm cr}\la10$ and that the fraction of
diffuse PeV neutrino flux contributed by blazars is $\la0.5$. These
are consistent with Ref. \cite{2014PhRvD..90b3007M} which suggests
that $\xi_{\rm cr}=30-300$ is required for blazars to account for
the IceCube detection, with the lower limit for the case of flat CR
spectrum.

\subsection{Caveats}
Some comments should be made here regarding the hypotheses in the
derivation.

First, we simply assume that the CR loading factor $\xicr$ is a
constant, which may not be true. There have been some research shows
that the radiative efficiency may be lower for higher luminosity
\citep{2010MNRAS.402..497G,2008MNRAS.387.1669G}, which implies that
$\xicr$ may weakly increase with $L_{\rm rad}$
\citep{2014PhRvD..90b3007M} and hence $L_\gamma$. The blazars in the
stacking searches of neutrinos by IceCube are mostly bright blazars,
i.e., with high gamma-ray luminosity and/or small distance, thus the
constraint to $\xicr$ (eq. \ref{eq:cr loading factor}) is mainly
valid to bright blazars. If the weak blazars are with smaller
$\xicr$ then the derived upper limit for the diffuse neutrino
intensity from blazars is even lower, since the main contribution
comes from lower luminosity and/or farther blazars. This leads to
even stronger conclusion that blazars cannot contribute to the
IceCube detection. The constraint is not valid if $\xicr$ decreases
strangely as the detected gamma-ray flux increases.

Second, the CR loading factor for BL Lacs may be different from that
of FSRQs. However it should be $\sim10^2$ times larger than
constrained by stacked FSRQs, in order for BL Lacs to account for
the IceCube flux. The constraint from 27 LSP BL Lacs,
$\xicr<0.92\zeta^{-1}$, does not agree so.

Third, some notes should be made here regarding the possible effects
of electromagnetic cascade on the estimate of the bolometric
radiation luminosity, and hence the cosmic ray luminosity. The
cascade may happen inside the sources, thus part of the high energy
gamma-rays is re-emitted in lower than Fermi-LAT energy range. Our
approach essentially takes the cascade into account, because the
cascade component will be a part of the bolometric radiation
luminosity.

On the other hand, the cascade may happen outside of the sources,
i.e., in the cosmic radiation background (CRB). However, the effect
can be neglected because: first, the 0.1-100 GeV gamma-rays hardly
suffer from absorption in the CRB, and the cascade basically ends at
$<100$ GeV; secondly, the IceCube measured neutrino flux, much
smaller than the all-sky integrated blazar 0.1-100 GeV gamma-ray
flux, implies that the cascade component is negligible compared to
the 0.1-100 GeV gamma-ray flux. Thus, the 0.1-100 GeV luminosity
provides a good estimate of the bolometric radiation luminosity. On
the contrary, if the cascade in propagation did happen seriously,
the bolometric radiation luminosity should be much larger than
estimated from the observed gamma-ray flux, then the constraint on
$\xicr$ by the same stacking limit for the neutrino flux should be
much more stringent, since the neutrino luminosity is proportional
to $\xicr L_{\rm rad}$. This does not weaken our constraints but
backward.

Finally, one may worry that there could be bias for the sample we
use because all these blazars are bright ones with high detected
gamma-ray flux. However, if the assumption that CR loading factor
$\xicr$ is roughly a constant holds then our constraint on $\xicr$
is valid no matter apparently bright or dim sample is used.

\subsection{Comments on recent works}
Our conclusion appears to be in conflict with some recent results.
It is recently reported \citep{2016arXiv160202012K} that a
high-fluence outburst of a FSRQ, PKS B1424-418, occurred in temporal
and positional coincidence with the third PeV neutrino event
detected by IceCube, indicating a direct physical association
between them. The about 1-yr outburst duration and the IceCube
effect area imply that the PeV neutrino flux is comparable to the
gamma-ray flux during the outburst, $S_\nu\sim
S_\gamma\sim10^{-7}\rm GeV\,cm^{-2}s^{-1}$
\citep{2016arXiv160202012K}. However, this seems to be in conflict
with two observational results: (1) as pointed out by
\cite{2016SCPMA..59a5759W}, the IceCube detected diffuse neutrino
flux (20TeV-2PeV) is only 4\% of the diffuse gamma-ray flux
(0.1-100GeV) from FSRQs; and (2) the stacking analysis by IceCube
for the catalog of 33 FSRQs even shows that the neutrino flux
\citep{2014ApJ...796..109A} is only $\la10^{-3}$ of their gamma-ray
one (see Table \ref{table:stacking}).

\begin{table}[t]
\begin{center}
\caption{Comparison of the total flux of gamma-rays and neutrinos
for the three blazar catalogs.}\label{table:stacking}
\begin{tabular}{lccc}
\tableline
Blazar catalog & $S_\gamma$ & $S_\gamma$  & $E_\nu^2 \Phi_{\nu_\mu + \bar{\nu}_\mu}^{90\%}$ \\
& 10-100GeV & 0.1-100GeV & $\sim$0.1-1PeV\\
\tableline
FSRQ  &   &  $4.08\times10^{-6}$ & $3.46\times10^{-9}$ \\
LSP BL Lac   & $8.41 \times10^{-9}$ & $1.31\times10^{-6}$ & $5.24\times10^{-9}$ \\
Hard-spectrum BL Lac   & $4.09\times10^{-8}$  & $2.76\times10^{-6}$ & $3.73\times10^{-9}$ \\
\tableline
\end{tabular}
\end{center}
Notes. All flux is in unit of  $\rm GeV \ cm^{-2} \ s^{-1}$. The
detailed lists of the three catalogs, 33 FSRQs, 27 LSP BL Lacs and
37 hard gamma-ray spectrum BL Lacs, are shown in Tables B.6, B.7 and
B.8, respectively, of \cite{Schatto2014}.
\end{table}

One may argue that the physical condition of the jet may change
with the outburst, especially the CR loading power. In the
outburst, with eq. (\ref{eq:individual flux}) one finds
$\xicr\sim8/f_{p\gamma}$, a factor $\ga8/0.062\sim130$ larger than
usual. Since $L_\gamma$ increases by 15-30 in the outburst
\citep{2016arXiv160202012K}, the CR luminosity ($L_{\rm
cr}\propto\xicr L_\gamma$) increases at least by (2-4)$\times10^3$'s
accordingly, which is a dramatically large contrast. PKS B1424-418 is indeed in the
FSRQ catalog for the track event search, with much larger effect
area than the HESE analysis at the source location, so it is
important to see if more PeV events can be found in the track-event
data during the outburst.

It is also suggested that for BL Lacs as a class to explain the PeV
neutrinos detected by IceCube, the neutrino flux from BL Lacs is
required to be comparable or larger than the gamma-ray flux above 10
GeV, $S_\nu\ga S_\gamma(>10\rm GeV)$ \citep{2015MNRAS.452.1877P}.
However, the comparison of Fermi-LAT and IceCube observations to two
catalogs of BL Lacs (Table
\ref{table:stacking}) shows that the BL Lac neutrino flux is, on the
contrary, smaller than their gamma-ray one.

A very latest investigation by IceCube collaboration
\citep{2016arXiv161103874I} contains the sample from the 2nd
Fermi-LAT AGN catalogue (2LAC). Still no excess is observed and
upper limits for cumulative neutrino flux of the resolved blazar
population is obtained. Only by assuming the proportionality of
gamma-ray and neutrino flux can they extrapolate the upper limit to
the all-sky blazars including the unresolved blazars in gamma-ray
observation. Their conclusion that 2LAC blazars contribute no more
than 50\% of the observed neutrinos if the spectral index is as hard
as $-2.2$ is consistent with ours.

\acknowledgements{This work is partly supported by the National
Natural Science Foundation of China under grant No. 11273005 and the
National Basic Research Program (973 Program) of China under grant
No. 2014CB845800.}

\end{document}